# Understanding Persistence: A 3D Trap Map of an H2RG Imaging Sensor

Rachel E. Anderson[1], Michael Regan[1], Jeff Valenti[1], and Eddie Bergeron[1]

*Space Telescope Science Institute*

**ABSTRACT**

*Several theories exist to explain persistence, most of which revolve around the distribution of traps. We aim to simulate persistence and illustrate this complex issue with a 3D trap map. For this experiment, we vary the detector voltage bias to simulate the change in the depletion region that occurs when the detector is exposed to light. This allows us to measure the distribution of traps in the depletion region. This paper will explore the results from this experiment and discuss the implications.*

## 1. INTRODUCTION

From an image processing perspective, persistence is the imprint of the previous image onto subsequent exposures that decays with time. Persistence occurs in IR detectors whenever a pixel is exposed to light, and can show up within the next exposure, or in an exposure hours later. The level of persistence not only depends on the accumulated charge, but the amount of time the charge sat on the detector and time since the last exposure. The current best method to mitigate this effect is to simply wait for the persistence to decay. Persistence affects scheduling, data reduction, and science, and it is a significant problem for the Hubble Space Telescope (HST) and the James Webb Space Telescope (JWST).

A simple model of the physical process behind persistence is shown in Figure 1-1, a modification of Figure 1 in Smith et al. (2008). Here, persistence is described as caused by a flash of light during an exposure (e.g. a cosmic ray) since it is simpler to depict than constant illumination (although the physics for constant illumination is the same [1]). On the left, we show the steady state at the start of an exposure, where all the traps in the deletion region are empty and all the traps in the free charge region are full. Following the start of the exposure there is a flash of light, causing a decrease in the depletion region. Now there are empty traps in the free charge region, which will cause negative persistence [2]. An example of this can be seen when looking at the slopes before and after a large cosmic-ray event where the slope after the cosmic-ray will be less than the slope before [2]. In regards to constant illumination, rather than a flash of light, as charge accumulates in a pixel more traps become accessible in the free charge region. If the detector sits in this state for several hours (not resetting) then we arrive at a steady-state, where again all traps in the depletion region are empty, and those in the free-charge region are filled. On the far right we show the state after a reset, where now the depletion region is increased (however, not back to its initial state), and now there are filled traps in the depletion region. This will cause positive persistence in the next exposure as these traps in the depletion region empty. Our diagram below differs from the Smith et al. (2008) model only in that we add trap capture to the picture.

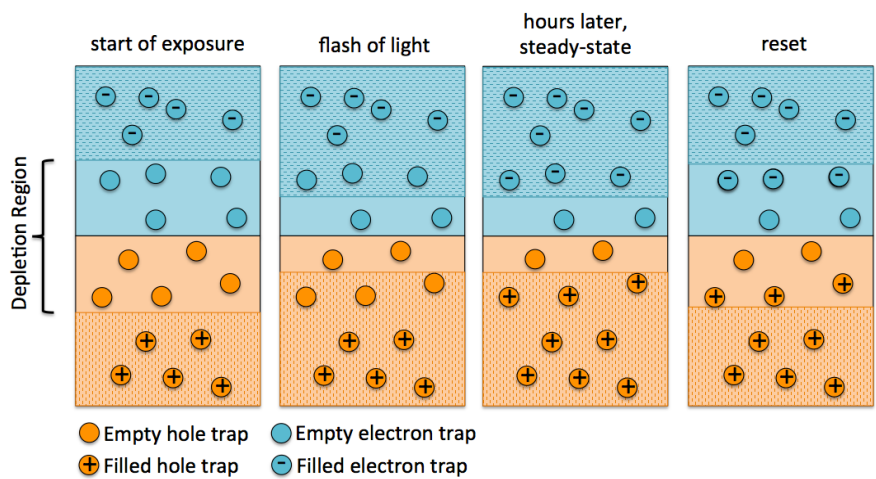

*Figure 1-1: A model of an infrared detector during the creation of persistence. In the first panel the detector is in a steady-state at the start of an exposure where all traps in the free-charge region are filled and all traps in the depletion region are empty. Next, a flash of light decreases the depletion region leaving empty traps in the free-charge region. Hours later the empty traps in the free-charge region have been filled causing negative persistence, and again the detector is in a steady state. Upon reset the depletion region increases, leaving filled traps in the depletion region. It is the decay of these traps that will cause positive persistence in the next exposure. Figure modified from Smith et al. (2008).*

Figure 1-1 displays persistence as an effect of illumination, but in fact the same story can be told using a change in bias, creating persistence electronically [1, 3]. The flash of light is equivalent to a decrease in the net bias, and the reset is equivalent to an increase in the net bias. There are several benefits to creating persistence electronically. First, it is possible to make precise incremental changes in the depletion region, as well as measure traps in a specific slice of the depletion region. Also, by decreasing the net bias you can demonstrate negative persistence, and by increasing the net bias show positive persistence [2]. Furthermore, it is simple to change the net bias and let the detector sit at that level until it reaches a steady-state by filling or emptying traps. Finally, persistence can be studied without external illumination. This technique is not only more precise, but may allow persistence to be characterized during I&T or during parallel calibrations on orbit. Therefore, in creating the 3D trap map, we create persistence electronically.

The simple picture shown in Figure 1-1 depicts persistence as a simple percentage of signal, as if we can assume uniform trap density. However, the geometry is more complicated than that.  The surface between the N-type and P-type materials are not cylindrical as in Figure 1-1, but curved [1].  Furthermore, defects at the surfaces are expected to cause an increase in the trap density at the surface [1].  If the defects are not uniform then the trap density will also not be uniform.  Therefore, as we change the size and location of the depletion region, the density of traps could change as well.  An example of this can be found in the Hubble Space Telescope's Wide Field Camera 3 (WFC3) data, where below about half saturation (~40,000 $e^-$ for WFC3) there is a very limited amount of persistence, but persistence grows rapidly above this point [4].  Therefore we know that there can be varying trap densities, and regions in the depletion region where there are little to no traps.

The goal of our study is to gather information to understand trap capture and release enough to model persistence as charge is accumulated during an exposure.  In order to do so, we need to know how many traps are filled as a function of location within the depletion region.  We determine the geometry of the trap density and create a 3D trap map by varying the bias voltage of an H2RG detector.  We proceed in steps, which is comparable to a short flashes of light rather than constant flux as noted above.  Note that a flash will show more persistence for the same final flux level due to the fact that the traps will be exposed to charge for longer and therefore capture more charge.  Two dimensional trap maps are simple to create with illumination, however we need electronic persistence to bring us to the third dimension, i.e. location within the depletion region.

## 2.   EXPERIMENT DESIGN

The data for this study was taken in the Operations Detector Laboratory (ODL) at the Space Telescope Science Institute on an engineering grade H2RG detector manufactured in 2005. From other experiments we know that this detector has a large variation in the trap density across the device, as shown by the 2D trap density map in Figure 2-1.  This 2D trap density map was creating by taking a high signal to noise dark after saturating the detector for about an hour.  Note that the low persistence regions (darker regions) are similar to expected JWST detector persistence levels.  Science grade detectors do not have these highest persistence pixels, but they are interesting here to help explore a larger variance in trap density.

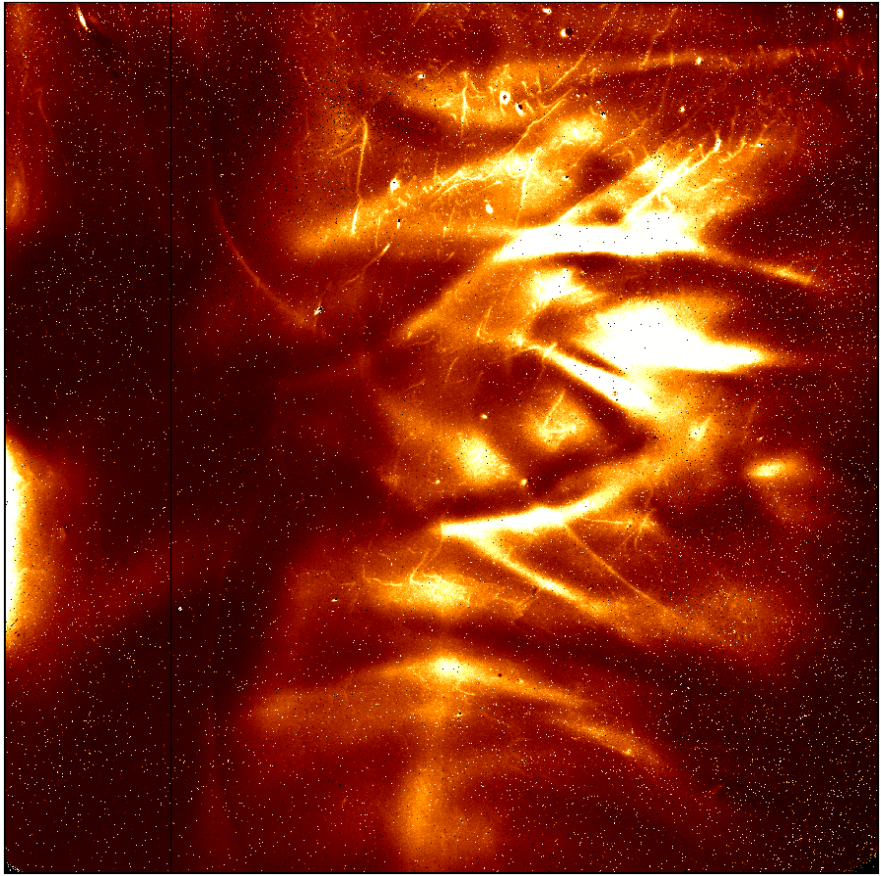

*Figure 2-1: 2D trap density map of H2RG detector. Brighter regions have more traps. The trap density per pixel varies by more than a factor of 20 across the array.*

The data was taken at several voltages on two separate dates, as shown in Table 2-1 below.

*Table 2-1: Data for this experiment was taken on two separate dates at varying voltages. On both dates we started with the largest voltage and stepped to the lowest voltage before stepping back up to the largest voltage.*

| Date | Substrate Voltages (mV) | Step Size (mV) |
|---|---|---|
| 11/17/2012 | 350 to 10 to 350 | 10 |
| 12/19/2012 | 120 to 0.0 to 120 | 10 |

At each voltage step, the following procedure was followed:
1. Change DSUB
2. Take six hours of darks to measure release / capture of traps. Each dark was a single integration of eight frames. In six hours we took ten darks with an integration time chosen to match the final dark, 2144.5 s.

3. Take high signal to noise final dark to measure dark current in a stable steady state. The final dark was a single integration of 200 frames, with a frame time of 10.72 s, and total integration time of 2144.5 s.

Only the final dark was taken with 200 frames in order to obtain a high signal to noise measurement of the final steady-state dark current. All data was taken at 37K, and a reset voltage of 85mV. Although these values were the commanded voltages and temperature, actual voltage and temperature information was captured in the headers and logs and used in the analysis. By maintaining a constant reset voltage (zero point), we only changed when the detector reached saturation. We chose this method instead of varying the reset voltage to simplify staying in the A-to-D range.

Results from the November dataset showed that we had stepped through the applied bias but not the natural depletion region. The purpose of the second run in December was to apply a forward bias to the detector to get closer to full well. The November and December data together cover the full voltage range of interest with a small overlap for substrate voltages 10mV to 120mV to verify that behavior in this range is consistent in the two separate voltage sweeps.

## 3. DATA REDUCTION

The following steps were taken for the reduction of this data:
1. Temperature correction was applied to each frame based on the temperature logs. This correction subtracted an ADU/K image scaled by the temperature difference between the current frame and the first frame.
2. We dark-subtracted the first dark after changing the DSUB from the final steady-state dark in order to show change in traps per ±10mV change in bias.
3. A reference pixel correction was applied to each dark-subtracted image. The correction subtracted the mean value of the reference pixels from the science pixels for each amplifier.

The corrected array was divided into 18 bins based on the 2D trap density map in Figure 2-1 after a cut was made to exclude outliers. Bins 1 – 9 are displayed in Figure 3-1 below, bin 1 having the lowest density of traps. Clearly bin 1 has the largest population, however, the minimum population per bin is 10,000 pixels, allowing us to maintain the signal to noise even in the highest density bins. The minimum bin width is 4410 traps/pixel.

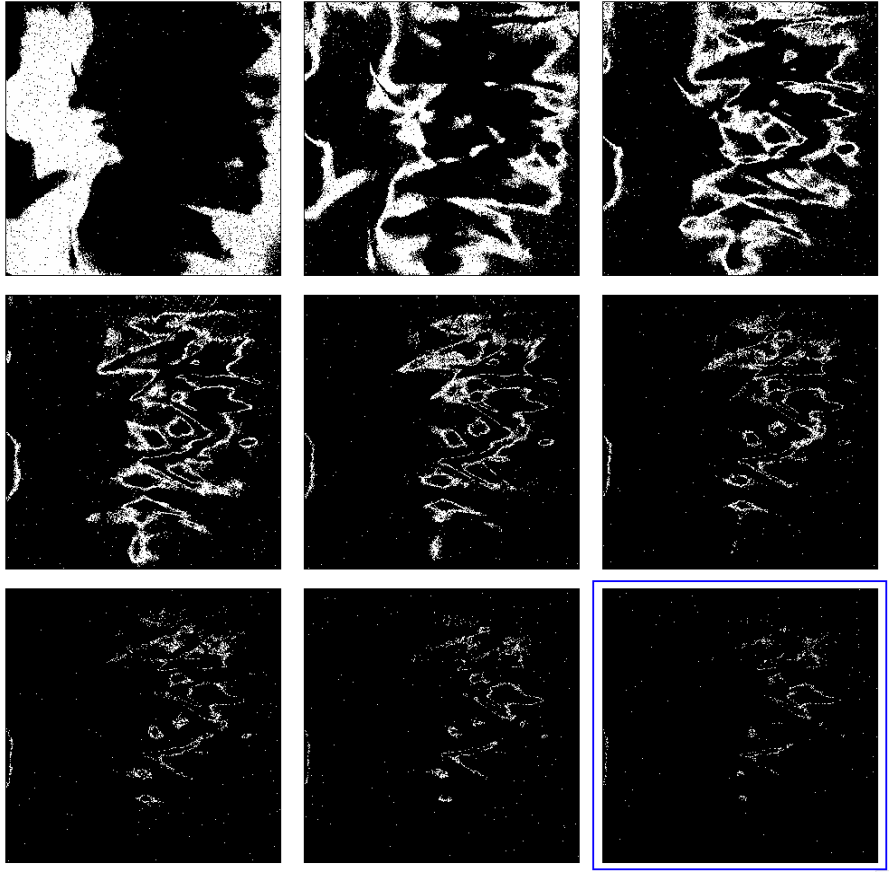

*Figure 3-1: Trap density bins 1 through 9 based on 2D trap density map. Bin 1, in the upper left corner, has the lowest density of traps and contains the most pixels. There are 18 bins total.*

## 4. RESULTS

The results from Figure 4-1 are consistent in the overlap region (from 10mV to 120mV) between the November and December datasets. We therefore average the two datasets together in the following figures. Furthermore, we started taking data immediately after the change of the substrate voltage, so the data marks the change going from one bias level to another. Therefore, for each of the plots below the bias voltage plotted is always the highest bias, regardless of whether the bias was increasing or decreasing.

The average 3D trap map per bin is displayed in Figure 4-1. Each line in Figure 4-1 is normalized by subtracting the lowest trap density bin, hence there are only 17 lines plotted. In Figure 4-1, we see negative persistence when decreasing the bias, and positive persistence when increasing the bias, as predicted. Furthermore, we observe a small linear change in trap density as we step through full well, with the trap density decreasing as we approach saturation. Despite the variance in the ODL H2RG detector (Figure 2-1), there is little structure, except at a bias of 137mV. Each point is an average of at least 10,000 samples (more than 1,300,000 for low persistence regions), and the structure is encountered when increasing and decreasing the bias. We therefore are seeing real structure in the trap density. With the WFC3 detectors as another example of non-uniform trap density, we conclude that a 3D trap map will be needed for each detector in order to model its persistence behavior.

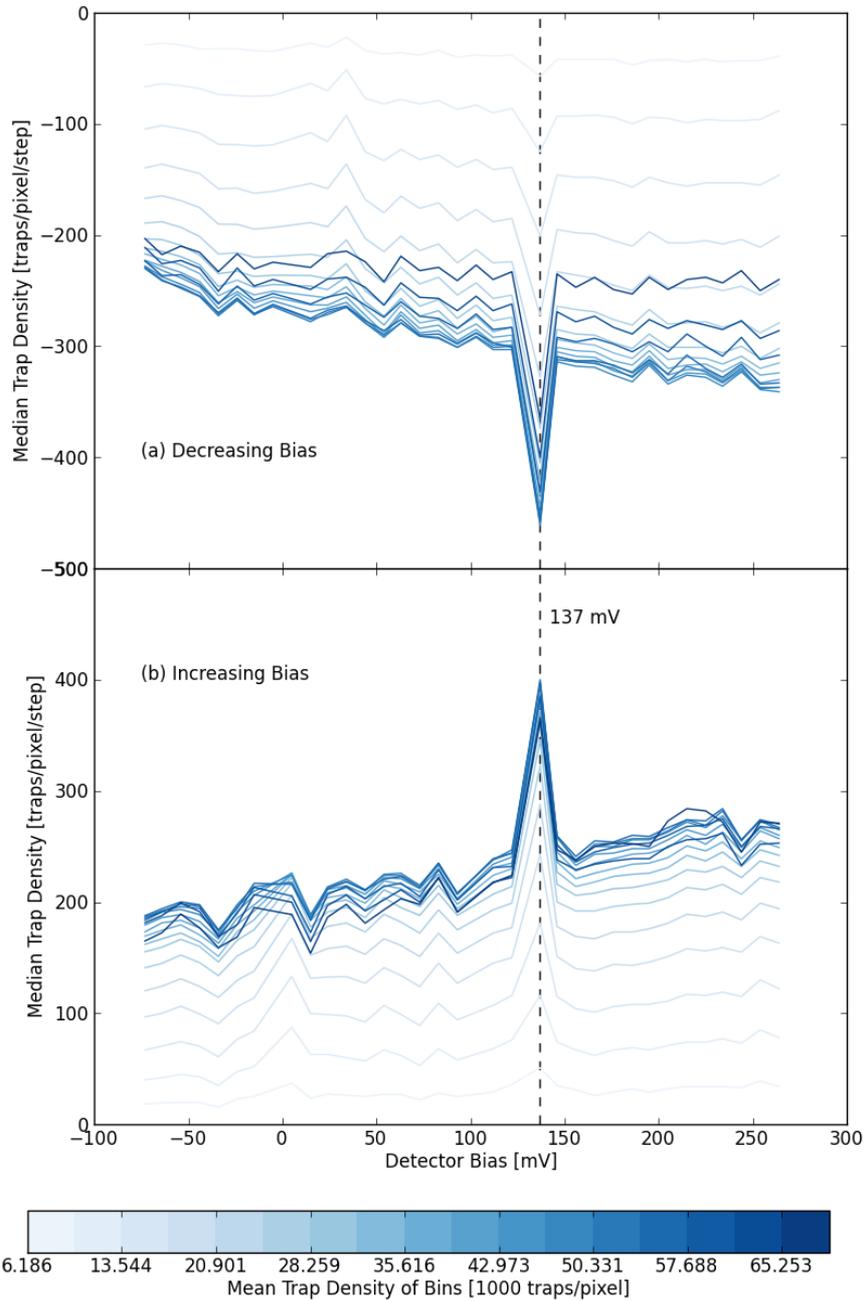

*Figure 4-1: The average 3D trap map per bin. The y-axis is the dark subtracted first dark after changing the DSUB.*

In Figure 4-2, we plot the change in the trap density as a function of the average value of each bin (count-rates from the 2D trap map) for both the increasing and decreasing bias cases. Note that we use the absolute value of the change in the trap density in order to make a better comparison between results obtained while increasing and decreasing the bias. Figure 4-2 shows that increasing the bias results in a shallower change in trap density per change in mV. We ignore the higher trap density bins in our conclusion of this plot, as they are unusual. Flight detectors do not have pixels such as these, and they are in the regime with short time-scale traps, etc. The difference we see between increasing and decreasing biases could be due to residual trap capture/decay even after we have waited six hours.

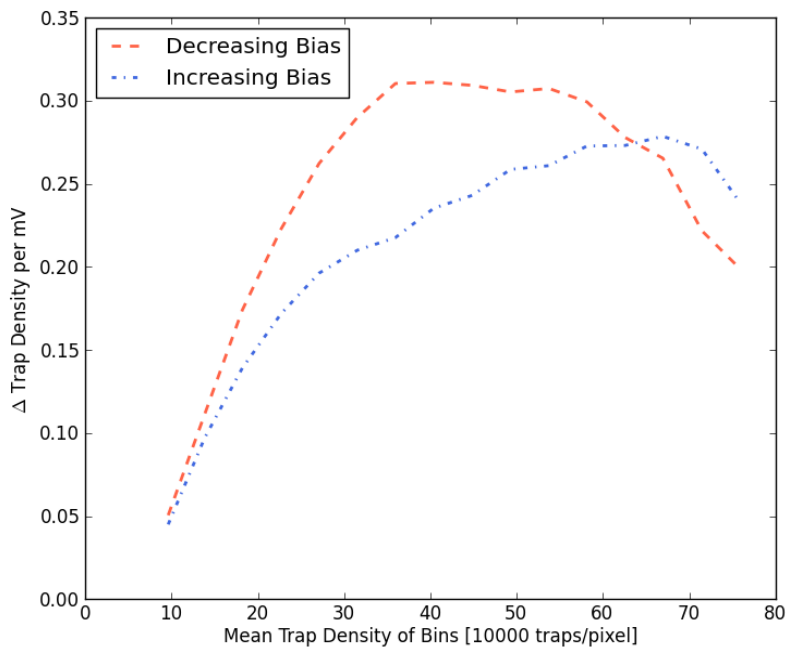

*Figure 4-2: Absolute change in the trap density as a function of bias vs. the average count rate of the 18 trap density bins.*

## 5. CONCLUSIONS

We have demonstrated the capabilities and benefits of introducing persistence electronically as well as a method for creating a 3D trap map of an infrared detector. The H2RG detector used here, as well as the HST WFC3 detector, shows evidence of structure in the trap density. Therefore, while it still might be possible to assume constant density for most detectors, 3D trap maps of more detectors are needed to make that conclusion. We find that our method as described in this report is well suited for mapping traps in 3D for infrared detectors.

Results from the 3D trap map of the ODL H2RG detector show a mostly linear relationship between persistence and the depletion region, although some structure is observed at a bias of 137mV.  Also, traps are seen throughout the depletion region, but decrease in density as we approach saturation.  That is to say increasing the bias increases the depletion region, in turn increasing persistence.  This is being looked into for the JWST Near Infrared Spectrograph (NIRSpec) detectors.

Our goal with this experiment is to be able to accurately model persistence. Creating a 3D trap map assists in this regard, but more information is still needed.  Future work includes understanding the time constants for both the decay time and filling of traps in order to include that in our model. We also need to compare these results to flight grade devices which have an significantly lower trap density.

## 6.    REFERENCES

[1]  R. M. Smith, M. Zavodny, G. Rahmer, M. Bonati, "A theory for image persistence in HgCdTe photodiodes", Proceedings of the SPIE, Vol. 7021, 2008.

[2]  M. Regan, E. Bergeron, K. Lindsay, R. Anderson, "Count rate nonlinearity in near infrared detectors: inverse persistence", Proceedings of the SPIE, Vol. 8442, 2012.

[3]  M. Regan, E. Bergeron, R. Anderson, "Maximizing Science from JWST Detectors", Proceedings of the Scientific Detector Workshop, this Volume, 2013.

[4]  K. S. Long, S. M. Baggett, J. W. MacKenty, A. G.. Riess, "Characterizing Persistence in the IR detector within the Wide Field Camera 3 Instrument on Hubble Space Telescope", Proceedings of the SPIE, Vol. 8442, 2012.